\documentclass[twocolumn]{article}          
\usepackage{amsmath}
\usepackage{graphicx}
\usepackage{mathptmx}      
\usepackage{float}
\usepackage[document]{ragged2e}
\usepackage{natbib}
\usepackage[hidelinks]{hyperref}
\usepackage{geometry}
\usepackage{hyperref}
\usepackage{xcolor}
\usepackage{subfig}

\usepackage{xcolor}
\hypersetup{
    colorlinks,
    linkcolor={red!50!black},
    citecolor={blue!50!black},
    urlcolor={blue!80!black}
}

\usepackage{multirow}
\usepackage[affil-it]{authblk}
\usepackage[english]{babel}
\usepackage[none]{hyphenat}
\usepackage{fancyhdr}
\title{Upstream Shear Layer Stabilisation via Self-Oscillating Trailing Edge Flaplets}
\renewcommand{\thefootnote}{\fnsymbol{footnote}}
\author[1]{Edward Talboys \thanks{Corresponding author: edward.talboys.1@city.ac.uk}}
\author[1]{Christoph Br\"{u}cker}
\affil[1]{City, University of London\\ Northampton Square, London, EC1V 0HB}
\date{}

\begin{document}
\maketitle
\renewcommand{\thefootnote}{\arabic{footnote}}
\thispagestyle{fancy}
\justify
\begin{abstract}
The flow around a symmetric aerofoil (NACA 0012) with an array of flexible flaplets attached to the trailing edge has been investigated at Reynolds numbers of 100,000 - 150,000 by using High-Speed Time-Resolved Particle Image Velocimetry (HS TR-PIV) and motion tracking of the flaplets' tips. Particular attention has been made on the upstream effect on the boundary layer evolution along the suction side of the wing, at angles of attack of 0$^\text{o}$ and 10$^\text{o}$. For the plain aerofoil, without flaplets, the boundary layer on the second half of the aerofoil shows the formation of rollers as the shear-layer rolls-up in the fundamental instability mode (linear state). Proper Orthogonal Decomposition (POD) analysis shows that non-linear modes are also present, the most dominant being the pairing of successive rollers. When the flaplets are attached, it is shown that the flow-induced oscillations of the flaplets are able to create a lock-in effect that stabilises the linear state of the shear layer, whilst delaying or damping the growth of non-linear modes. It is hypothesised that the modified trailing edge is beneficial for reducing drag and can reduce aeroacoustic noise production in the lower frequency band, as indicated by an initial acoustic investigation.

\end{abstract}

\begin{table}[H]
\centering
\begin{tabular}{lll}
\hline\noalign{\smallskip}
Symbol & Units & Description \\
\noalign{\smallskip}\hline\noalign{\smallskip}
U$_\infty$  &  m/s     & Free-stream velocity     \\
U$_0$& m/s & Convective velocity\\
U$_{\text{es}}$& m/s & Boundary layer edge velocity\\
$u'$&  m/s & Streamwise velocity fluctuations\\
$v'$& m/s & Wall normal velocity fluctuations\\
c & m & Aerofoil chord\\
Re$_{\text{c}}$& --- & Chord based Reynolds Number\\
$\alpha$& $^\text{o}$ & Angle of attack\\
$f_0$& Hz & Shear layer fundamental frequency\\
$f_{0-1/2}$& Hz & Shear layer instability frequency\\
$f_k$& Hz & Vibration frequency at k$^\text{th}$ mode\\
$\alpha_k$& --- & Wave-number at the k$^\text{th}$ vibration mode\\
$\delta^*$ & m & Boundary layer displacement thickness\\
$\lambda_0$& m & Shear layer roll-up wavelength\\
$\text{St}^*_0$& --- & Shear layer Strouhal Number\\
s& m & Spanwise width of flaplet\\
L& m & Length of flaplet\\
d& m & Spacing between flaplet\\
h& m & Thickness of flaplet\\
w& m & Flaplet vertical displacement\\
D & kg/s & Damping coefficient\\
$\zeta$ & --- & Damping ratio\\
$\omega_n$ & rad/s & Natural Frequency\\
$\omega_d$ & rad/s & Damped Frequency\\
$\phi$ & rad & Phase delay\\
E& N/m$^\text{2}$& Elastic Modulus \\
I& m$^\text{4}$& Area moment of inertia \\
$\rho_f$ &kg/m$^\text{3}$ & Flaplet material density\\
K & --- & Non-dimensional bending stiffness\\
\noalign{\smallskip}\hline              
\end{tabular}
\caption{Nomenclature Table}
\label{Table: Nomenclature}
\end{table}


\section{Introduction}
\label{sec: Intro}
In recent years extensive research has been carried out by turning to nature for inspiring new designs in man-made aircraft and improving aerodynamic performance or reducing emissions such as noise or pollution. Ever increasing air travel and human habitation near airports, has led to the urgent need for cleaner and quieter air travel. A few solutions which have been previously investigated are; leading edge undulations \citep{Miklosovic2004}, serrated trailing edges \citep{Howe1991}, slitted trailing edges \citep{Gruber2010}, brush-like edge extensions \citep{Herr2005} and porous aerofoils \citep{Geyer2010}. In the present study, a flexible trailing edge consisting of an array of small elastic flaplets, mimicking the tips of bird feathers, is used. 

The specific arrangement and structure of feathers on the trailing edge of an owl's wing is well known to be one of the key mechanisms that the bird uses to enhance noise suppression \citep{Jaworski2013}. In addition, secondary feathers on the upper side of the wings of the steppe eagle (\textit{Aquila nipalensis}) \citep{Carruthers2007} and the peregrine falcon (\textit{Falco peregrinus}) \citep{Ponitz2014}, to give a few examples, have been observed to pop-up as the birds attack prey or come into land at a high angle of attack. This phenomenon has been the subject of many research studies.

\citet{Schluter2010} could show that by attaching rigid flaps, via a hinge, on the upper surface of the wing, the  C$_\text{L}$ max is increased for a series of tested aerofoils (NACA 0012, NACA 4412, SD 8020). Schl\"{u}ter also showed that the flaps bring  the additional benefit of gradual stall rather than a more severe lift crisis. \citet{Osterberg2017} carried out a similar study on a flat plate subjected to high angles of attack, coming to the same conclusion.

\citet{Brucker2014} used flexible flaplets attached on the suction side of a NACA 0020 aerofoil that was subjected to a ramp-up motion. The results show a considerable delay in dynamic stall. Flap length and flap chordwise spacing were varied and it was found that the most successful configuration was two rows of flaps of length 0.1c spaced 0.15c-0.2c in the chordwise direction. The wavelength of the rollers in the shear layer was found to be of the same order, 0.15c-0.2c. This led to the conclusion that the flaplets resulted in a lock-in effect with these rollers, as the two spacial length scales were comparable. Concluding that this lock-in effect stabilises the shear layer. \citet{Rosti2017} then built on these findings of by carrying out a DNS (direct numerical simulation) parametric study. The flap element was rigid but coupled to the aerofoil surface by a torsional spring type coupling. It was found that when the flap oscillations were at the same frequency as the shear-layer roll up, the mean lift coefficient was at its highest. 
\begin{figure*}[ht!]
\centering
{\includegraphics[width = 1\linewidth]{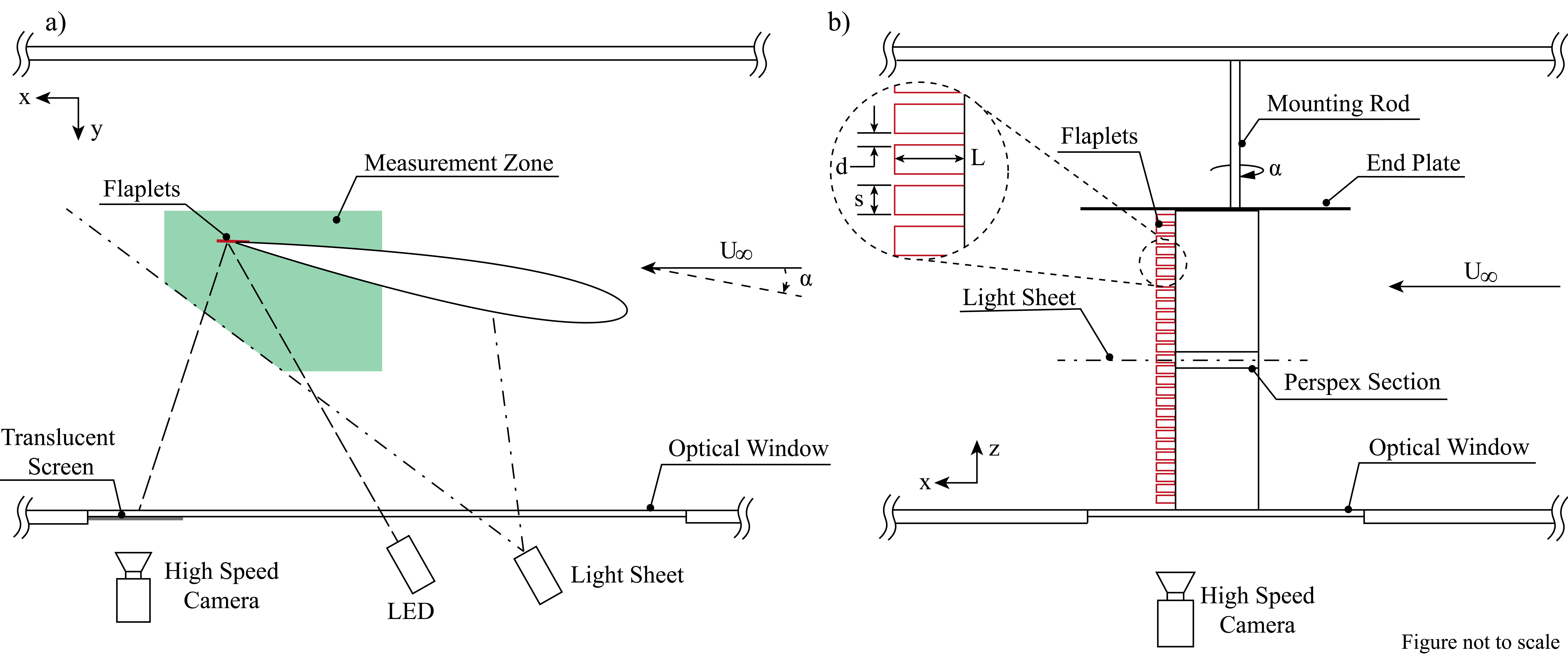}}
\caption{a) Plan view of the experimental set-up. b) Side view of the experimental set-up}
\label{fig: Set-Up}
\end{figure*}

Similar effects, as observed for aerofoils, have also be seen on tests with bluff bodies. A series of studies on cylinder flows with flaplets attached on the aft half of the cylinder have been carried out by \citet{Kunze2012, Kamps2017a} and \citet{Geyer2017}. \citet{Kunze2012} saw that the presence of the flexible elements allowed the shedding frequency to be locked in with the most dominant eigen-frequency of the flaplets. This hints to a similar lock-in effect of the shear layer roll-up observed for the aerofoils, \citet{Rosti2017}. Consequently the flow-structure in the wake is changed and a reduced wake deficit is observed. \citet{Kamps2017a} then tested the same cylinder/flaplet configuration in an aeroacoustic wind tunnel and could show that the flaplets overall reduce noise, both in the tonal and the broadband components.

\citet{Talboys2018a} have very recently, during the review process of the current study, carried out an aeroacoustic campaign with the present configuration of aerofoil and self-oscillating trailing edge flaplets. They have found that when the flaplets are attached to the trailing edge, an accoustic reduction in the low to medium frequency range (100~Hz - 1~kHz) is observed over a large range of Reynolds Numbers and angles of attack. This supports the findings of the preliminary investigation carried out by \citet{Kamps2017} on a NACA 0010 with flexible flaplets attached to the trailing edge. In another recent publication \citet{Jodin2018} uses an active control technique with a vibrating solid trailing edge, which can oscillate up to 400~Hz with a peak deflection in the order of 1~mm. By using high-speed PIV in the wake it was seen that when the trailing edge oscillations are actuated, the wake thickness was reduced by as much as 10\% and accompanying lift force measurements showed a 2\% increase.   

The present study builds off these previous studies in order to study the benefit/impact of flexible flaplets being attached to the trailing edge rather than on the aerofoil body. Tests of such modifications of the trailing edge by \citet{Kamps2017} and \citet{Talboys2018a} already show promising results in noise reduction. However, no details of the flow structure and the fluid-structure interaction are known which might explain the observed aeroacoustical modification. This is the motivation and purpose of the present study.

\section{Experimental Set-Up}
\label{sec: Exp Arr}
The experiments were carried out in the Handley Page laboratory at City, University of London in a closed loop wind tunnel. The test section of the tunnel is 0.81~m by 1.22~m in cross-section, and has a turbulence intensity of 0.8\%. A NACA 0012 aerofoil, with a chord of 0.2~m and span of 0.52~m, was used for the present study. One side of the aerofoil spanned to the floor of the tunnel and an endplate was affixed to the exposed end to negate any end effects. The aerofoil was 3D printed in two sections with a small perspex section in the measurement window, as per Figure \ref{fig: Set-Up}b. The use of perspex in this region improves the quality of the PIV recordings close to the surface. Air flow with density of $\rho$ = 1.2~kg/m$^\text{3}$ was applied at different flow speeds $\text{U}_\infty$. Three chord based Reynolds numbers were analysed in this study: 100,000 125,000 and 150,000; at two different angles of attack: 0$^\text{o}$ and 10$^\text{o}$. A 0.3~mm thick boundary layer trip was implemented at 0.2c on both sides of the aerofoil to ensure that the boundary layer (BL) was turbulent.

The trailing edge with the flexible flaplets was made from a polyester foil, of thickness h = 180~$\mu$m, which was laser-cut at one long side to form an array of individual uniform flaplets. Each rectangular flaplet has on the long side a length of L = 20~mm and on the short side a width of s = 5~mm with a gap of d = 1~mm in spanwise direction (see \ref{fig: Set-Up}b). The foil was adhered to the pressure side of the aerofoil using thin double sided tape such that the flaplets face downstream, with their free end located at a distance of x/c = 1.1c downstream of the trailing edge. The flaplets form a mechanical system of a one-sided clamped rectangular cantilever beam which is free to oscillate perpendicular to the mean-flow direction. 

\subsection{Velocity Field Measurements}    
\label{sec: Exp Arr PIV}
High-Speed Time-Resolved Particle Image Velocimetry (HS TR-PIV) measurements were carried out using a 2~mm thick double pulsed Nd:YLF laser sheet in a standard planar set-up. A high speed camera (Phantom Miro M310, window size 1280 x 800 pixels) equipped with a macro lens, Tokina 100~mm, with f/8 was used in frame straddling mode. Olive oil seeding particles, of approximate size 1~$\mu$m, were added to the flow downstream of the model. A number of 500 pairs of images were captured at a frequency of 1500~Hz with the pulse separation time being altered for each case, given in Table \ref{Table: PIV Set-Up}. 

\begin{table}
\centering
\begin{tabular}{cccc}
\hline\noalign{\smallskip}
Re$_\text{c}$ & Aperture Size & Pulse Separation & Capture Frequency \\
\noalign{\smallskip}\hline\noalign{\smallskip}
100,000         & f/8      & 80~$\mu$s        & 1500~Hz\\
150,000         & f/8      & 30~$\mu$s        & 1500~Hz\\
\noalign{\smallskip}\hline              
\end{tabular}
\caption{TR-PIV parameters used for the two different Re$_\text{c}$}
\label{Table: PIV Set-Up}
\end{table}

The raw images were then processed using the TSI Insight 4G software which uses the method of 2D cross correlation. The first pass interrogation window size was 32 x 32 pixels, with a 50\% overlap. The size was then reduced to 16 x 16 pixels for the subsequent pass. A 3 x 3 median filter was then applied to validate the local vectors, any missing or spurious vectors were interpolated by using the local mean.

\subsection{Flap Motion Tracking}
\label{sec: Exp Arr FMT}
To track and record the real time motion of the flaplets with high resolution, a high power LED (HARDsoft IL-106G) was used alongside a second high speed camera (Phantom Miro M310) with a Nikon 50~mm f/1.8 lens. The LED was directed at the flaplets and the back reflections were seen on a translucent screen adhered to the transparent wind tunnel side wall (see Figure \ref{fig: Set-Up}a). Due to the optical lever-arm condition, small deflections of the flaplets led to a large displacement of the back-scattered light on the screen. The recordings were taken at 3200~Hz, with an exposure time of 320~$\mu$s and an aperture of f/1.8. In total 2.5 seconds of motion was recorded on the Phantom Camera Control Application (PCC), resulting in 8000 images for each case. An edge detection code (Matlab) was then used to track the flap tips motion over time. Spurious data points were removed from the data set and subsequently interpolated prior to a low pass filter (4th order Butterworth low-pass with a cut-off frequency of 600~Hz) being applied to the data.        
 
\section{Results}
\label{sec:Results}
The test cases mentioned in section \ref{sec: Exp Arr}, were run both with and without the flaplets in order to ascertain a baseline for comparative analysis. For the remainder of the report the nomenclature in Table \ref{Table: TM} will be used and only results for Re$_\text{c}$=100,000 and Re$_\text{c}$=150,000 are shown.  

\begin{table}
\centering
\begin{tabular}{cccc}
\hline\noalign{\smallskip}
Test Name & Re$_\text{c}$ & $\alpha$  & Trailing Edge \\
\noalign{\smallskip}\hline\noalign{\smallskip}
100-0-P         & 100,000      & 0         & Plain			\\
100-0-F         & 100,000      & 0         & Flaplets		\\
100-10-P        & 100,000      & 10        & Plain			\\
100-10-F        & 100,000      & 10        & Flaplets		\vspace{2mm}\\
150-0-P         & 150,000      & 0         & Plain			\\
150-0-F         & 150,000      & 0         & Flaplets		\\
150-10-P        & 150,000      & 10        & Plain			\\
150-10-F        & 150,000      & 10        & Flaplets		\\
\noalign{\smallskip}\hline              
\end{tabular}
\caption{Test matrix nomenclature}
\label{Table: TM}
\end{table}

\subsection{Flaplet response to step input}	\label{sec: Results Flap Eigen}
\begin{figure*}
\centering
{\includegraphics[width=1 \linewidth]{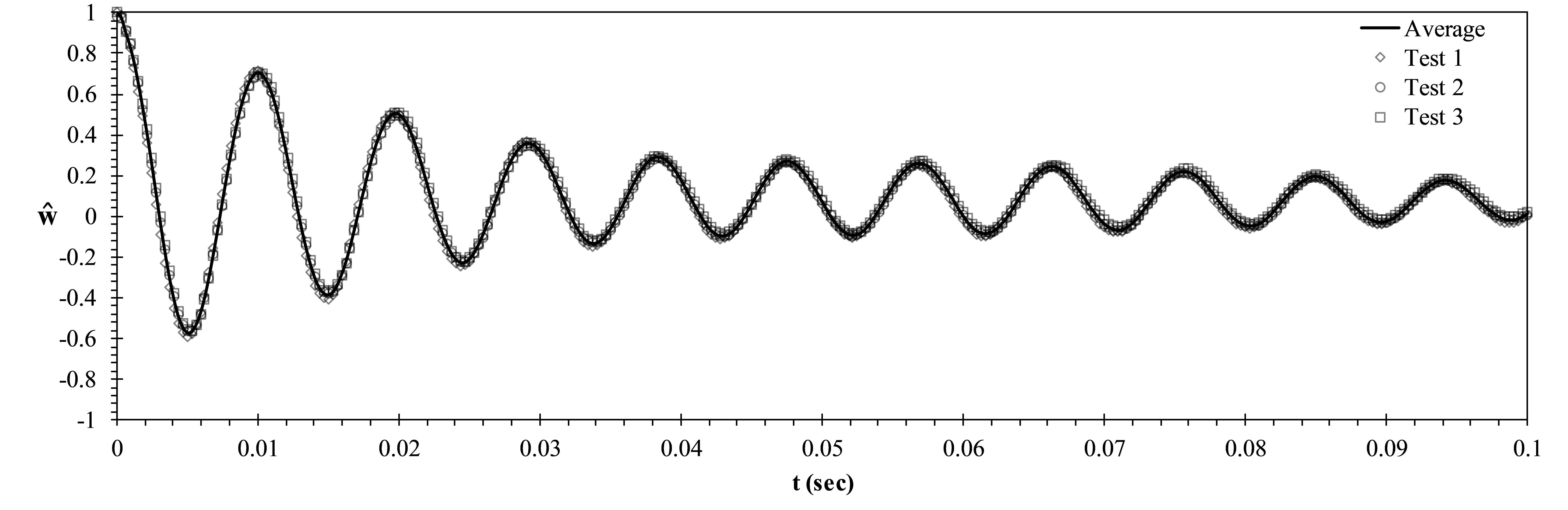}}
\caption{Response of a singular flaplet to a step input.}
\label{fig: StepResp.}
\end{figure*}

As previously mentioned (sect. \ref{sec: Exp Arr FMT}), the flaplets can be understood as thin rectangular cantilever beams, clamped at one side and free to oscillate in their natural bending mode in the flow environment. The subsequent equation of motion from bending beam theory leads to the following ODE of the mechanical system \citep{Stanek1965}.:  
\begin{eqnarray}
EI \frac{\partial^2 w}{\partial x^4} = -\rho_f A \frac{\partial^2 w}{\partial t^2} - \frac{D}{L} \frac{\partial w}{\partial t} \label{eqn: Cantilever Theory}
\end{eqnarray}
The general solution for a damped harmonic oscillators is then:
\begin{eqnarray}
\hat{w}(t) = \frac{w}{w_0} e^{-\zeta \omega_n t} cos(\omega_d t - \phi)\label{eqn: Underdamped System}
\end{eqnarray} 
In the case that the system is only weakly damped, the natural frequency results to:
\begin{eqnarray}
f_k = \frac{\alpha_k^2h}{2 \pi L^2} \sqrt{\frac{EI}{12\cdot\rho_f}} \label{eqn: Flap freq.}
\end{eqnarray} 

A step-response test was conducted with a single flaplet being bent out of the equilibrium to an amplitude of $w_0$, and subsequently unloaded, after which the flaplet oscillates back to its equilibrium position at rest. The tip motion was recorded by a high speed camera (Phamtom Miro M310, 1200 x 800~pixels at 3200~Hz) and the previously mentioned edge detection script was used to track the tip. The recorded response, Figure \ref{fig: StepResp.}, gives a Q-factor of 61.6 ($\text{Q}_{\text{factor}} = 1\text{/2}\zeta$), which indicates a very weakly damped harmonic oscillator. The natural frequency of the flaplets is 107~Hz as obtained by analysing the spectrum of the signal. With the known solution of equation \ref{eqn: Flap freq.}, for the weakly damped harmonic oscillator, and by using the first natural bending mode of the beam, $\alpha_1 = 1.875$, one can estimate the Young's Modulus of the flap material. Equation \ref{eqn: Flap freq.} has been further used to evaluate at which point the flaplets will go into the second vibration mode, for this case $\alpha_2 = 4.694$ leading to $f_2 = 671$ Hz. 

\begin{table}[h]
\centering
\begin{tabular}{cc}
\hline\noalign{\smallskip}
Property & Value \\
\noalign{\smallskip}\hline\noalign{\smallskip}
$f_1$ at $\alpha_1$		& 107~Hz						\\
$f_2$ at $\alpha_2$		& 671~Hz						\\	
$\rho_f$	& 1440~kg/m$^\text{3}$			\\	
$E$		& 3.12~GPa						\\	
\noalign{\smallskip}\hline              
\end{tabular}
\caption{Flaplet dynamics and material properties}
\label{Table: Flaplet Dynamics Data}
\end{table}

The non-dimensional bending stiffness of the flaplets has been calculated from equation \ref{eqn: Non-dim stiffness} giving a minimum value of K = 6.13 x10$^{-3}$ at Re$_\text{c}$ = 150,000 and a maximum, K = 13.91 x10$^{-3}$ at Re$_\text{c}$ = 100,000. Overall, it is concluded that the flaplets are of sufficient flexibility in order to easily react with pressure fluctuations in the present flow conditions. 

\begin{eqnarray}
\text{K} = \frac{EI}{\rho \text{L}^3 \text{U}_\infty^2} = \frac{E\text{s}}{12}\left(\frac{\text{h}}{\text{L}}\right)^3 \frac{1}{\rho \text{U}_\infty^2} \label{eqn: Non-dim stiffness}
\end{eqnarray}

\subsection{Flow Field}
\label{sec:PIV Results}

\begin{figure}
\centering
{\includegraphics[]{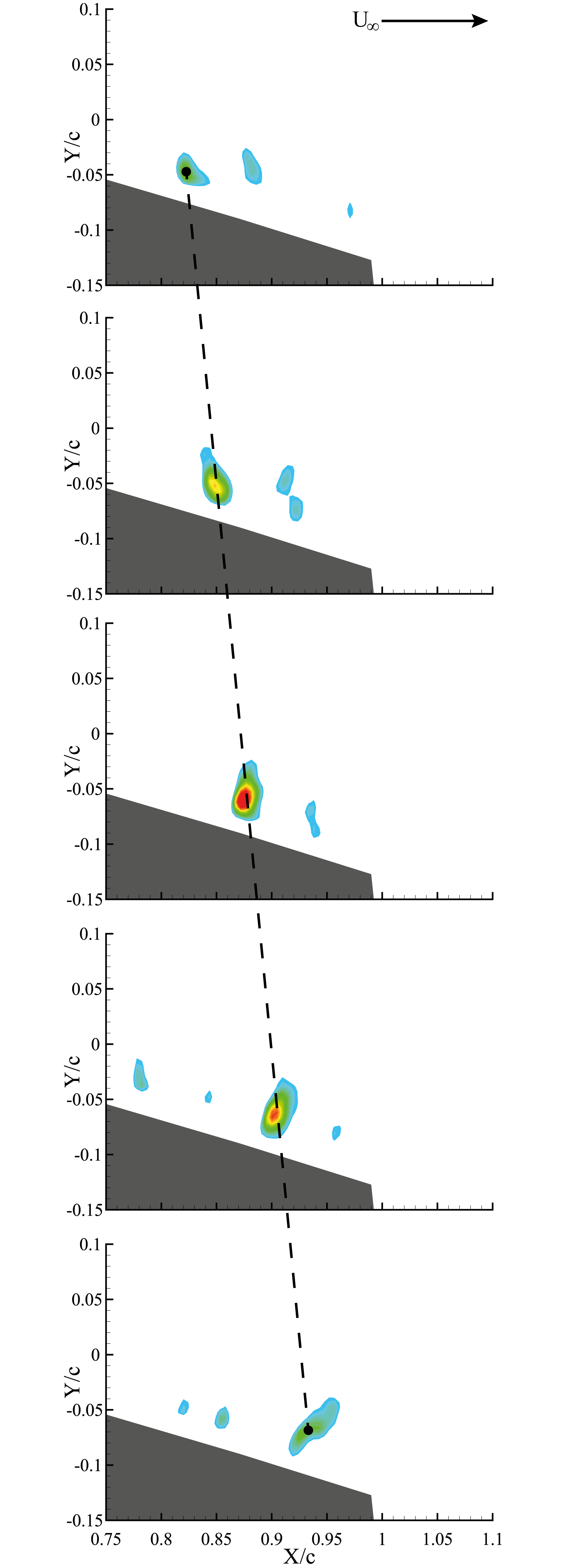}}
\caption{Q-Criterion for case 100-10-F. Time spacing in between each image is: $\Delta t^* = \Delta t\cdot U_\infty /c = 0.055$}
\label{fig: PIV Q Crit}
\end{figure}

The high-speed PIV measurements were analysed to gather the coherent structures in the shear layer by conditional averaging and POD analysis. Therefore, a virtual probe location has been selected within the PIV field, which acts to give an indicative value for the passage of rollers. It is found that the proper location in the field is where the maximum RMS streamwise velocity fluctuations ($u'$ RMS) occur. The wall normal position of the maximum $u'$ RMS gives an approximate boundary layer displacement thickness ($\delta^*$) and the position at which the inflection in the boundary layer occurs. Hence this location corresponds to the wall normal coordinate where the shear layer instability is most prevalent \citep{Dovgal1994}. For the cases 150-10-P and 150-10-F, $\delta^*$/c was found to be 0.027 and 0.028 respectively, at the chordwise location x/c = 0.85. Accordingly, the probe location has been set at x/c = 0.85 and y/c = $\delta^*/c$.

The fluctuating vertical velocity ($v'$) signal was recorded at the probe location (average of the 3x3 neighbouring vectors) and it was observed, from the PIV results, that the presence of a shear layer roller travelling through the probe area corresponded to a peak in the $v'$ component. Therefore in order to visualise a clear depiction of these structures passing through the probe area, the velocity fields were conditionally averaged with the peaks in $v'$. In order to observe how the structures develop in time, 5 frames preceding and following were also stored and averaged. From these averaged frames the Q-criterion was subsequently computed and the convection speed was calculated, see Figure \ref{fig: PIV Q Crit}. All cases fell in the range $\text{U}_0/\text{U}_\text{infty}$ = 0.45 -. A further breakdown of the individual cases can be seen in Table \ref{Table: Freq. Analysis}.

\begin{figure}[t!]
\begin{minipage}{1 \linewidth}
\centering
\subfloat[Mode 2, corresponding to $f_{0-1/2}$]{\label{POD:a}\includegraphics[scale=1]{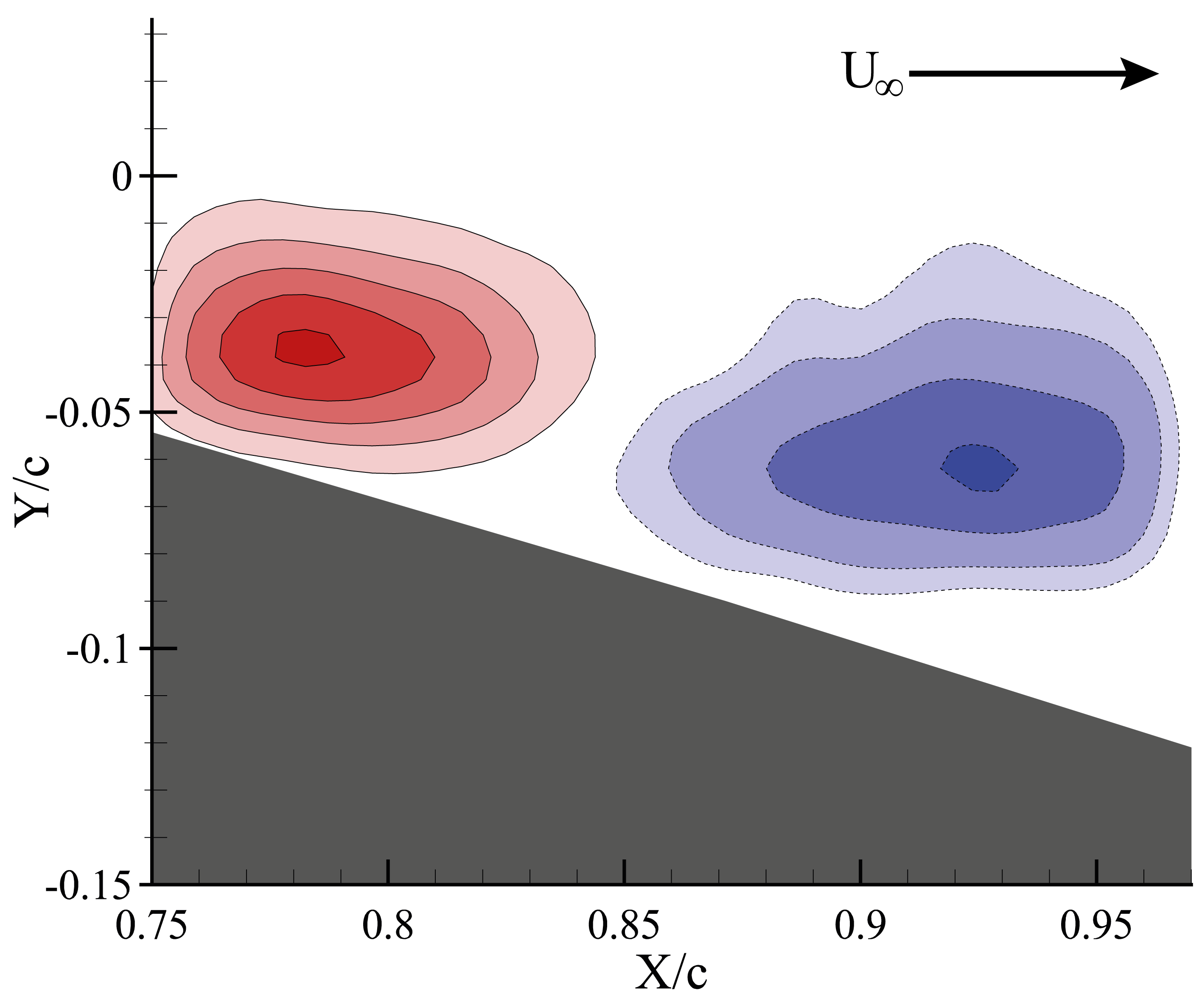}}
\end{minipage}\par \medskip
\begin{minipage}{1 \linewidth}
\centering
\subfloat[Mode 4, corresponding to $f_{0}$]{\label{POD:b}\includegraphics[scale=1]{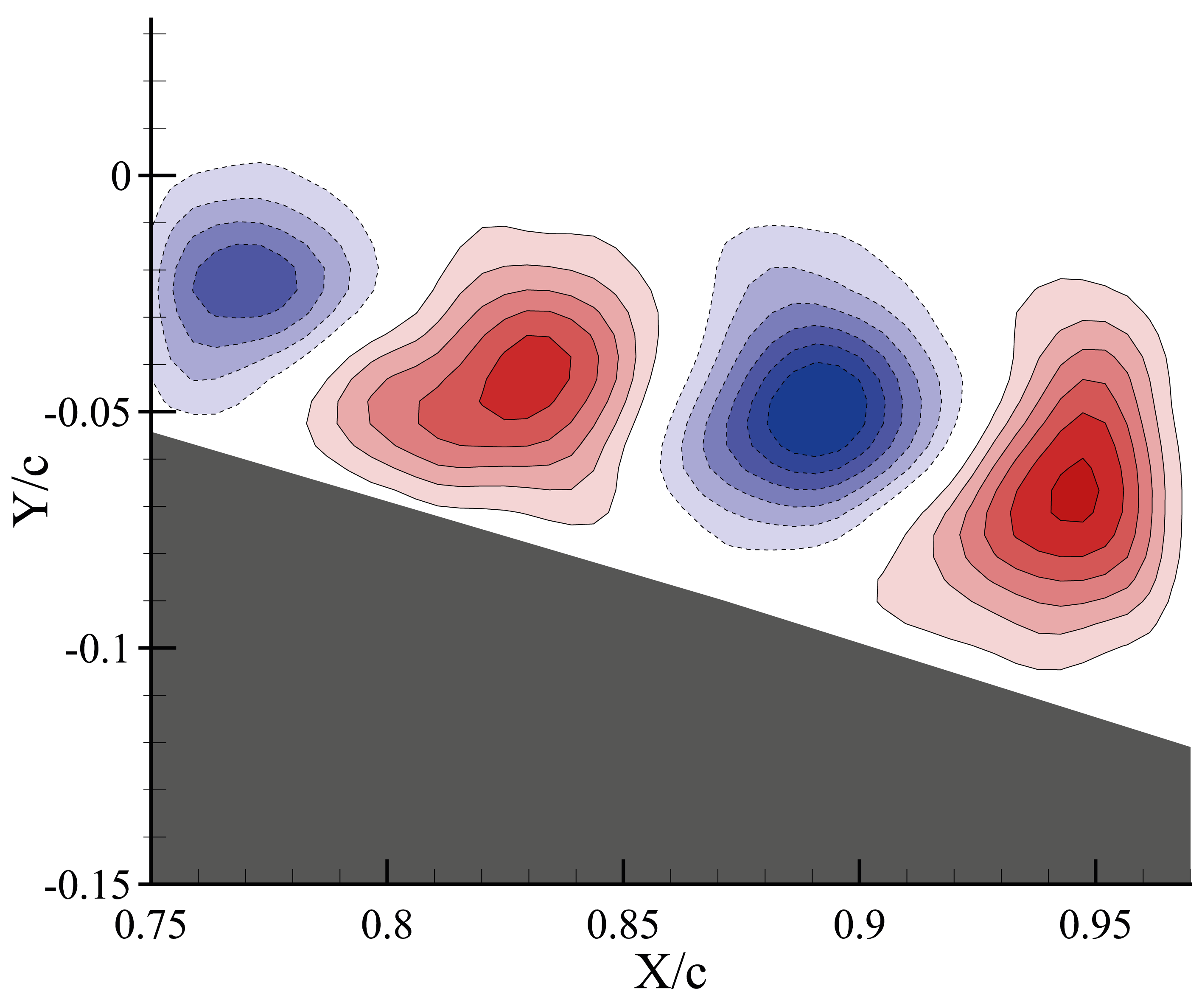}}
\end{minipage}\par
\caption{$v'$ POD mode comparison for the case 150-10-F.}
\label{fig: POD - Re150 10}
\end{figure}

In order to select the dominant modes in the flow field, a proper orthogonal decomposition (POD) of the fluctuating velocity field was carried out. POD reduces the field into modes, whereby each mode is a certain dominant flow feature or structure. Mode 1 represents the distribution of the Reynolds stresses in the boundary layer and is not shown here. In the present case the dominant features are the observed rollers that originate from the instability of the shear layer and the roll-up into vortices as seen in Figure \ref{fig: PIV Q Crit}. The spatial reconstruction of two of the dominant modes can be seen in Figure \ref{fig: POD - Re150 10}. Mode 4 (Figure \ref{POD:b}) is the mode which corresponds to the fundamental instability, leading to the shear-layer rolling up into regular rows of spanwise rollers. A proof of being it mode 4 is given later based on previous research of boundary layers on aerofoils. Then, mode 2, Figure \ref{POD:a}, corresponds to half of the fundamental instability which indicates the presence of a non-linear state in the shear-layer. This mode occurs at half the wavelength of the fundamental mode and represents a subharmonic feature generated by pairing of successive rollers. Mode 3 is a mixture of mode 2 and mode 4 and is linked  with the fundamental mode with a factor of 2/3. Thus it seems to be an intermediate state.       

\begin{figure*}[t!]
\begin{minipage}{.33 \linewidth}
\centering
\subfloat[150-10-P; PIV Probe]{\label{Spectra:a}\includegraphics[scale=.33]{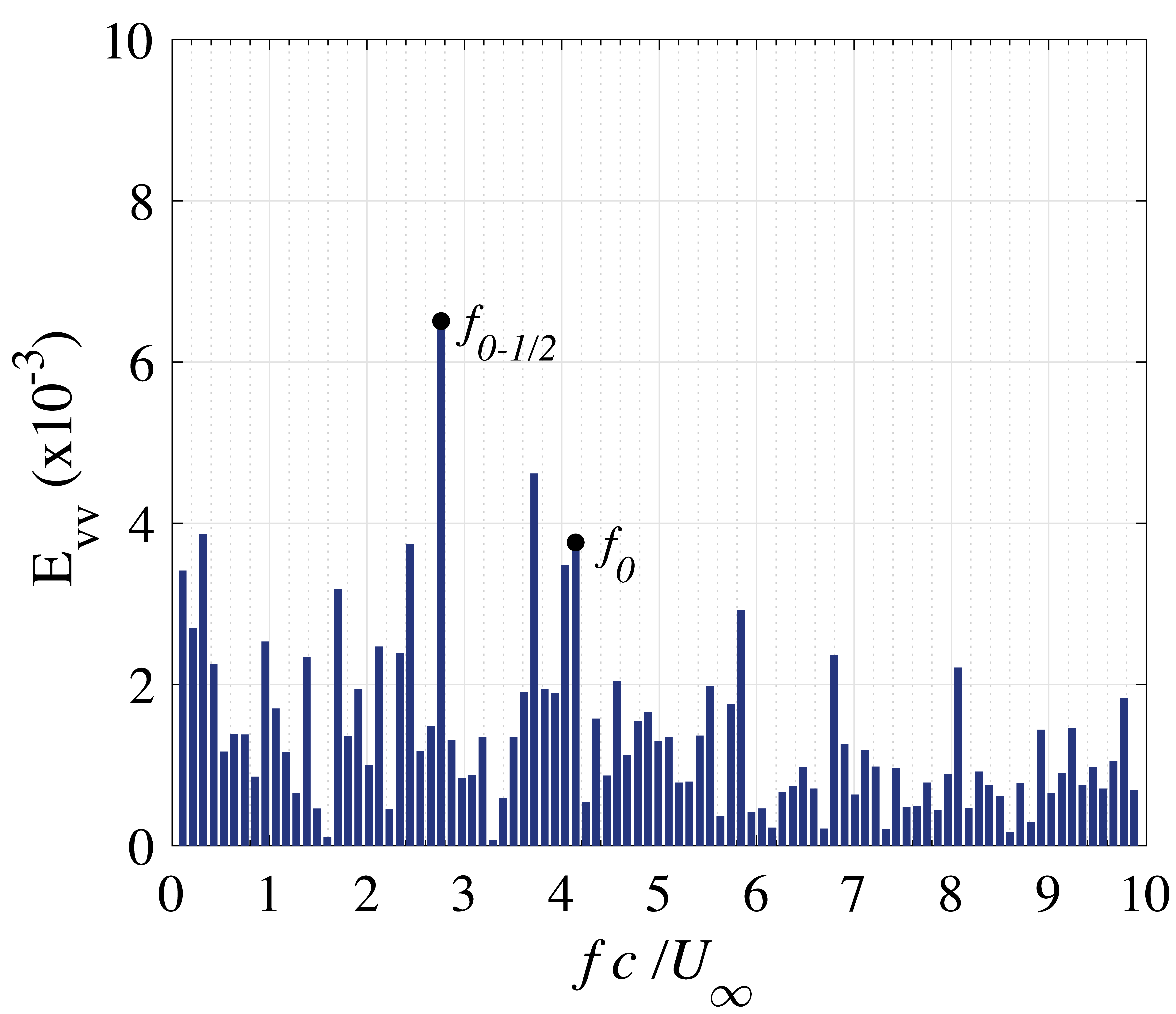}}
\end{minipage}
\begin{minipage}{.33 \linewidth}
\centering
\subfloat[150-10-P; 2$^\text{nd} v'$ POD mode]{\label{Spectra:b}\includegraphics[scale=.33]{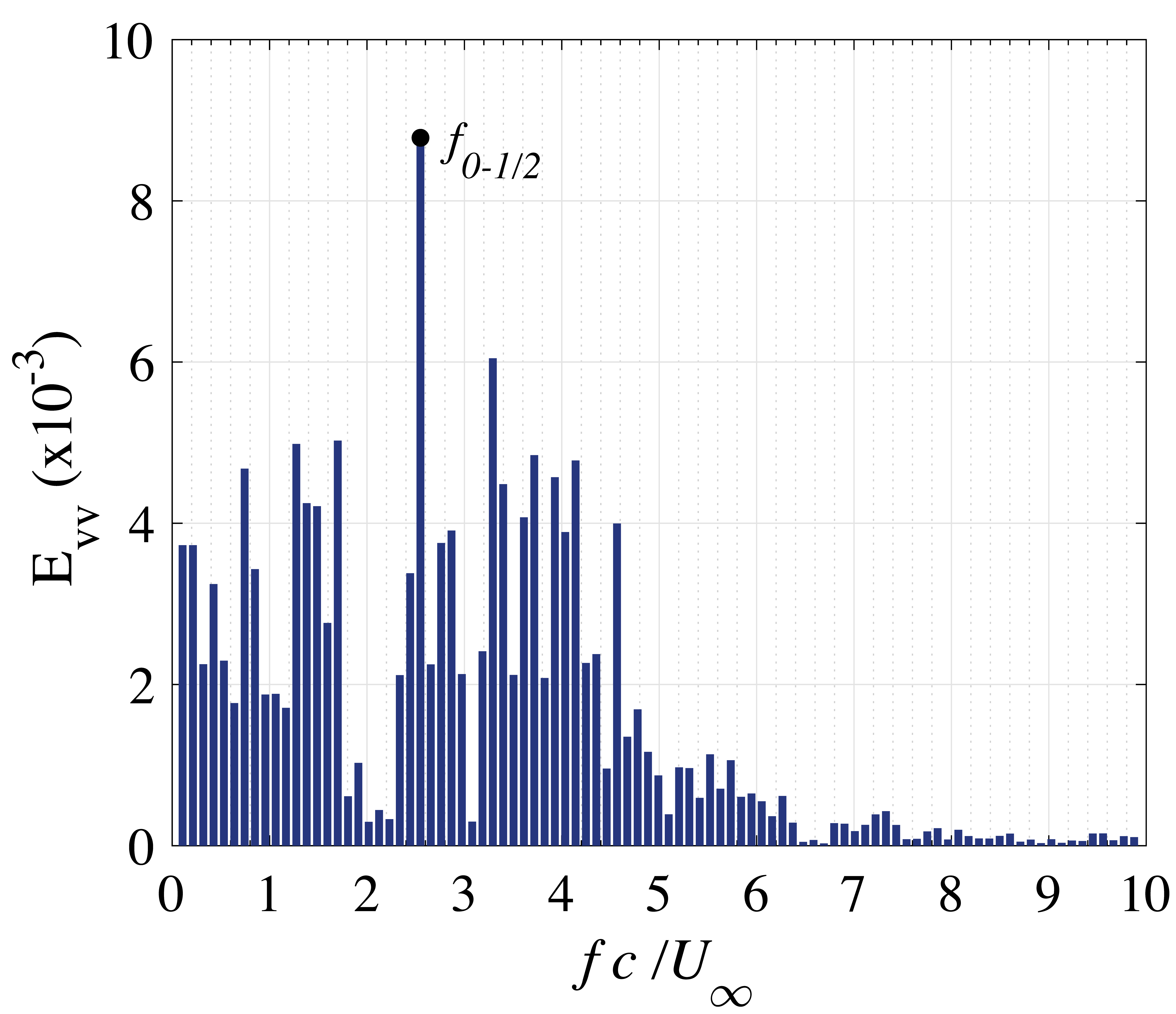}}
\end{minipage}
\begin{minipage}{.33 \linewidth}
\centering
\subfloat[150-10-P; 4$^\text{th} v'$ POD mode]{\label{Spectra:c}\includegraphics[scale=.33]{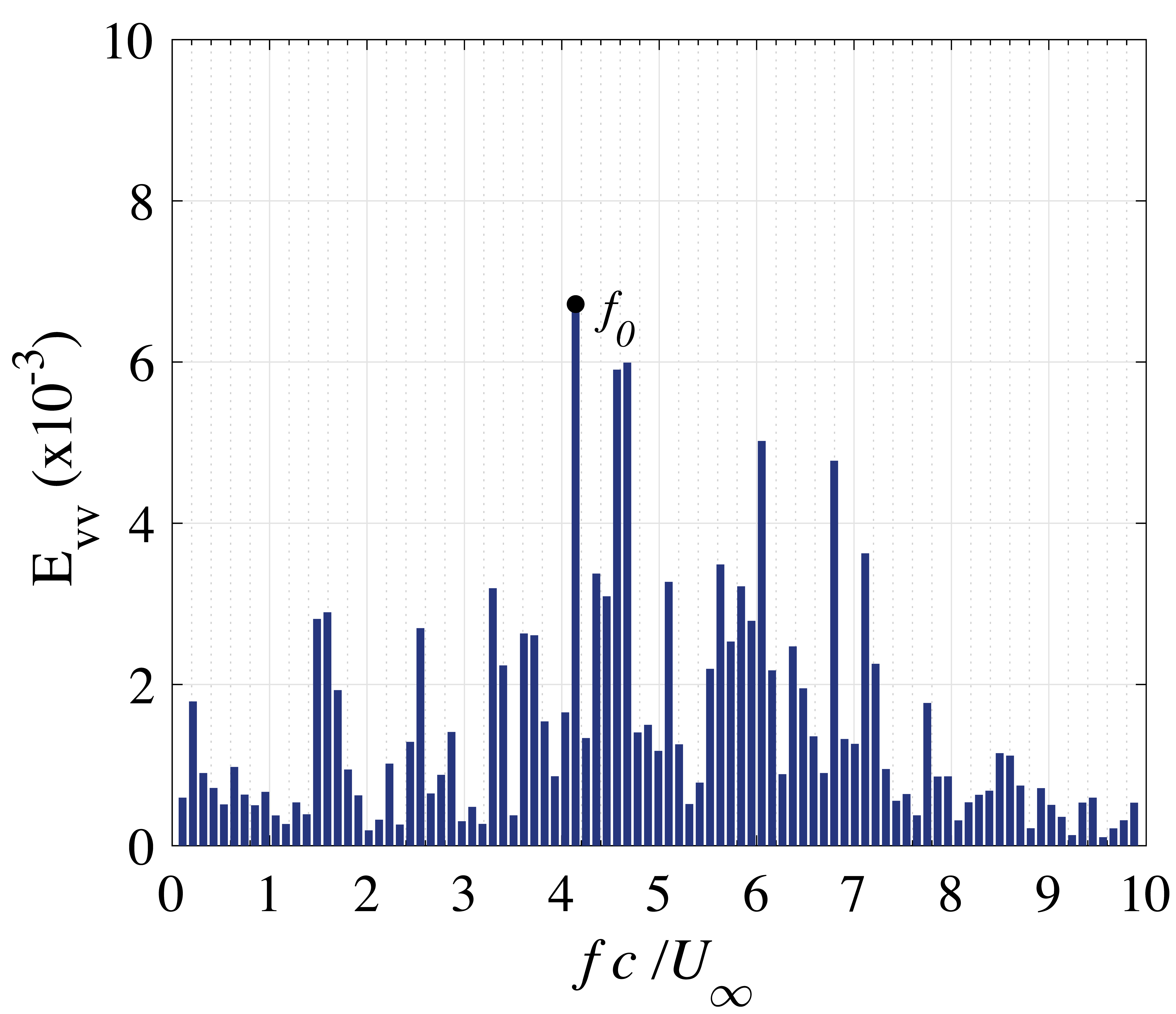}}
\end{minipage}\par \medskip
\begin{minipage}{.33 \linewidth}
\centering
\subfloat[150-10-F; PIV Probe]{\label{Spectra:d}\includegraphics[scale=.33]{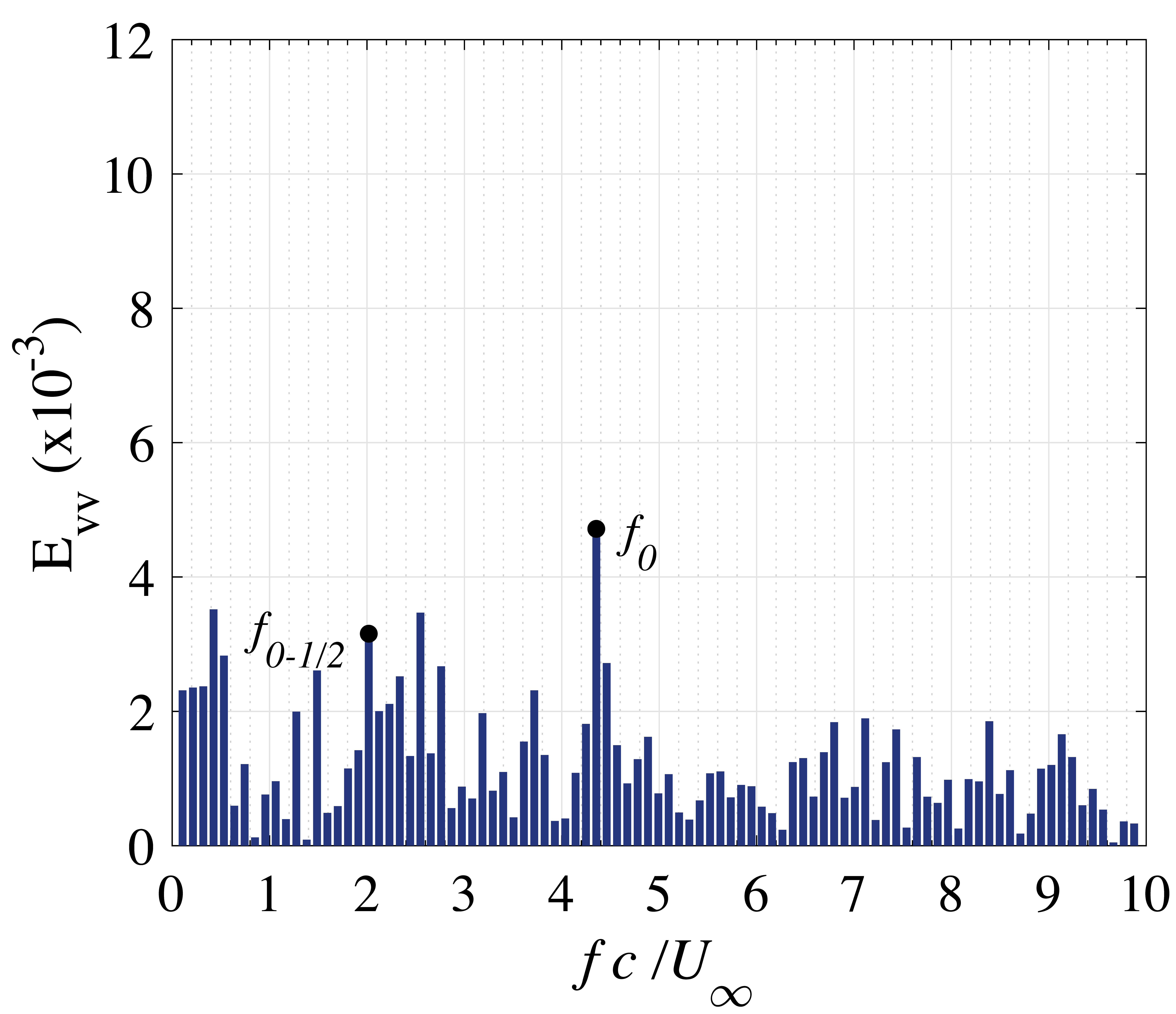}}
\end{minipage}
\begin{minipage}{.33 \linewidth}
\centering
\subfloat[150-10-F; 2$^\text{nd} v'$ POD mode]{\label{Spectra:e}\includegraphics[scale=.33]{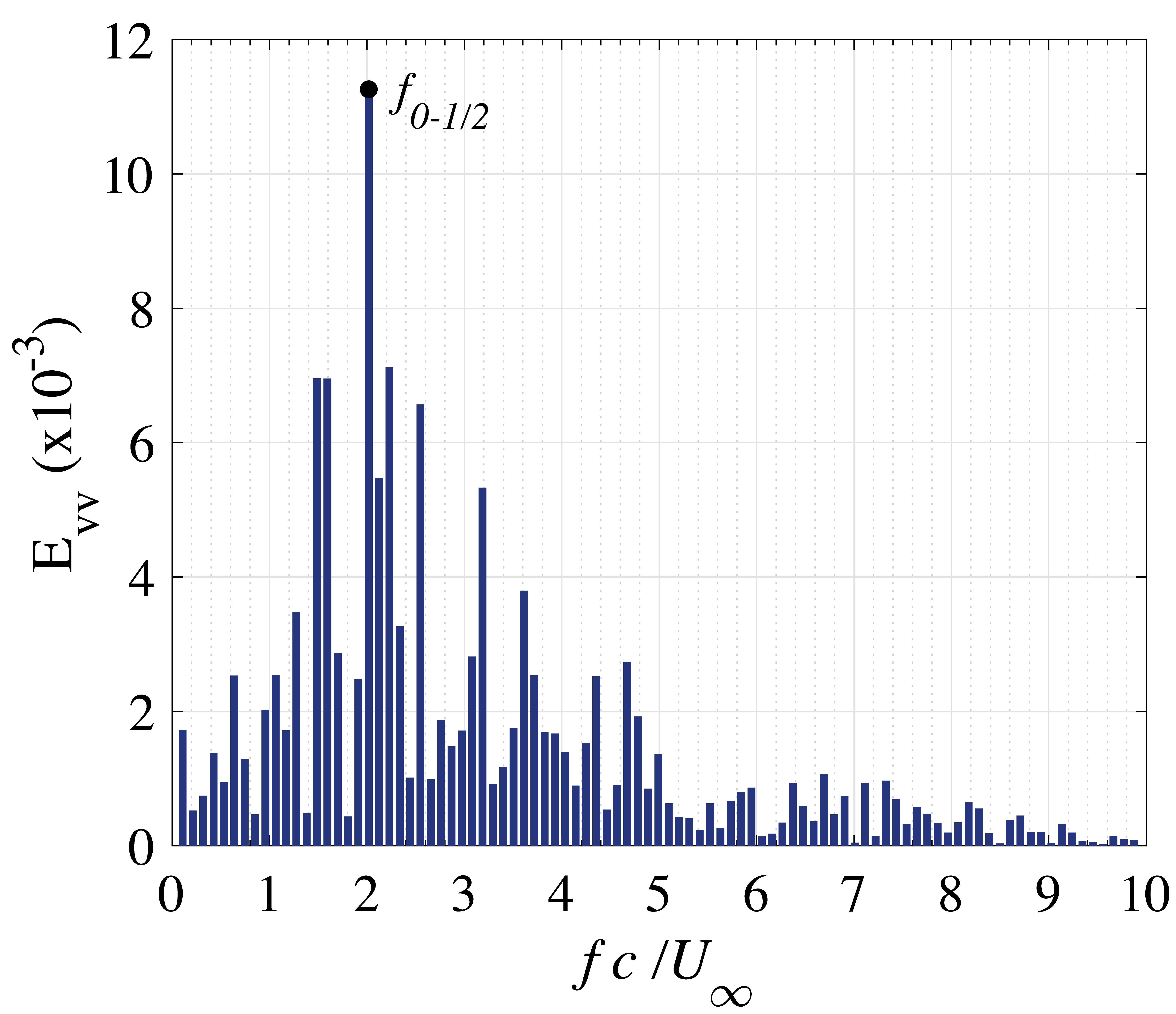}}
\end{minipage}
\begin{minipage}{.33 \linewidth}
\centering
\subfloat[150-10-F; 4$^\text{th} v'$ POD mode]{\label{Spectra:f}\includegraphics[scale=.33]{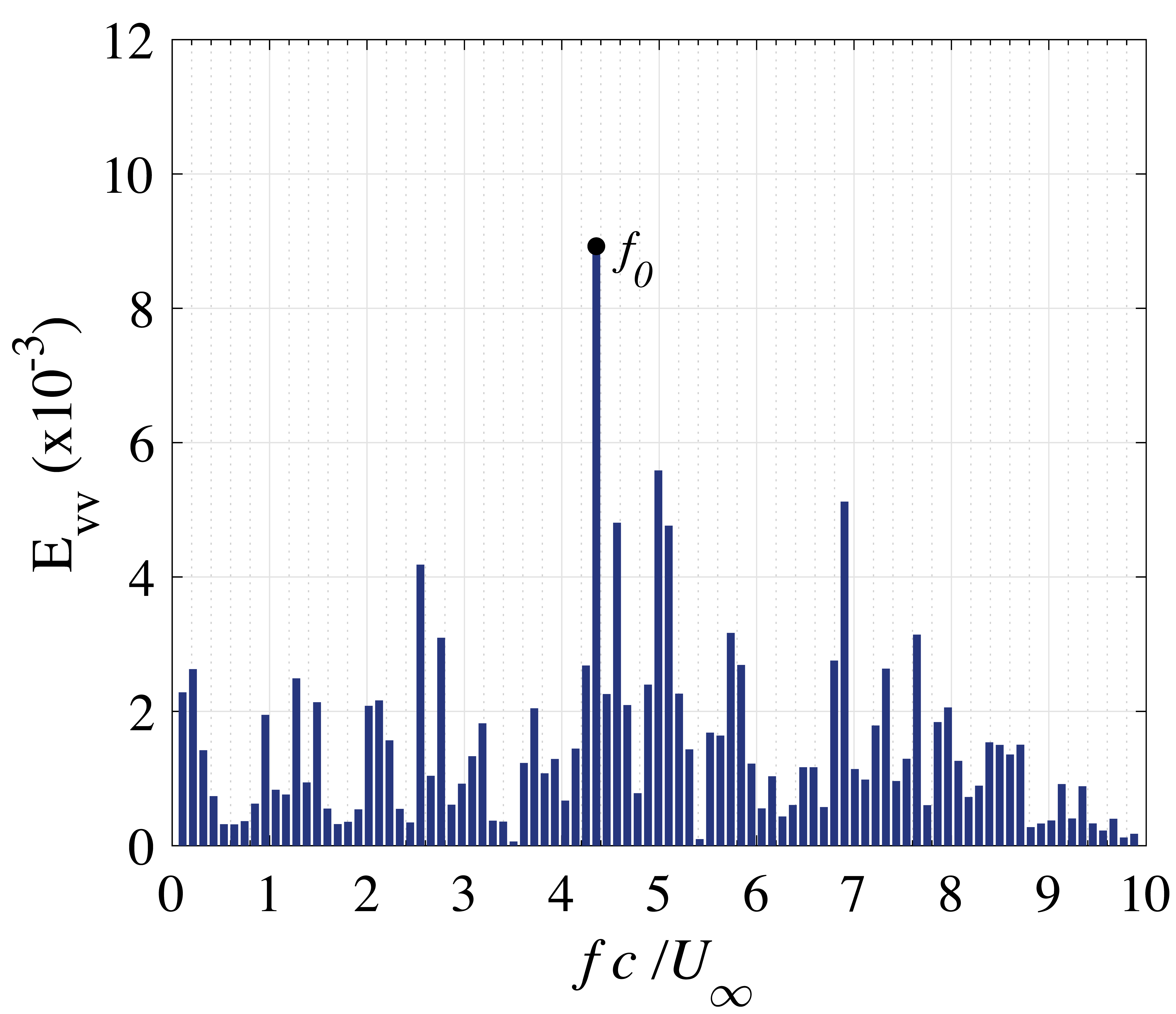}}
\end{minipage}\par
\caption{Spectral analysis of the $v'$ component from the velocity probe, 4$^\text{th} v'$ POD mode and 2$^\text{nd} v'$ POD mode at Re$_\text{c}$ = 150,000 and at $\alpha$ = 10$^\text{o}$. a-b-c: Baseline case. d-e-f: Flaplet case}
\label{fig: 150-10-Spectra}
\end{figure*}

In order to justify our selection of the mode 4 representing the fundamental shear layer instability, the Strouhal number of the mode was calculated and compared to previous research on similar aerofoils, see \citet{YARUSEVYCH2009}. They suggested a certain range of the Strouhal number exists in which the fundamental mode should fall in when scaling the frequency ($f_0$) with the wavelength of the fundamental roll-up ($\lambda_0$) and the boundary layer edge velocity ($U_{es}$), see equation \ref{eqn: Strouhal}. Analytically \citet{YARUSEVYCH2009} showed that this quantity should be in the region of 0.45 $\leq$ St$_0^*$ $\leq$ 0.5 and more recently this range was increased to 0.3 $\leq$ St$_0^*$ $\leq$ 0.5 by \citet{Boutilier2012}. Values for $\lambda_0$ were obtained from the POD (Figure \ref{fig: POD - Re150 10}) and the convection velocity of the vortex cores were calculated from the conditionally averaging, Figure \ref{fig: PIV Q Crit}. The values of St$^*_0$ obtained in the present study are within the limits set by \citet{Boutilier2012} and are in agreement also with previous studies \citep{Boutilier2012, Thomareis2017, YARUSEVYCH2009, Brucker2014}. Similar roller wavelengths were observed by \citet{Brucker2014}, where the wavelength they found was 0.15c - 0.2c. The value of the fundamental frequency could then be back calculated from this relationship and is presented in Table \ref{Table: Freq. Analysis} as $f_{0c}$. These calculated values show good agreement with the values obtained from the $v'$ spectra at the PIV probe location (Figure \ref{fig: 150-10-Spectra}), giving cofidence in the presented frequencies. Additional analysis is carried out on the temporal coefficients of the POD modes as they represent the temporal signature of the mode \citep{Semeraro2012, Meyer2007}. This helps to link the corresponding mode 4 with  $f_0$ and mode 2 with $f_{0-1/2}$, as the identified frequencies in the $v'$ spectra. 

\begin{table*}[h!]
\centering
\small
\begin{tabular}{c|cc|cc|ccccc|cc}
\hline\noalign{\smallskip}
\multirow{2}{*}{Test Cases} & \multicolumn{2}{c|}{PIV Probe} & \multicolumn{2}{c|}{POD} & \multicolumn{5}{c}{Shear layer Strouhal number parameters} & \multicolumn{2}{c}{Boundary layer}\\
 & \multicolumn{1}{c}{$f_0$ c/U$_\infty$} & \multicolumn{1}{c|}{$f_{0-1/2}$ c/U$_\infty$} & \multicolumn{1}{c}{$f_0$ c/U$_\infty$} & \multicolumn{1}{c|}{$f_{0-1/2}$ c/U$_\infty$}& \multicolumn{1}{c}{U$_0$/U$_\infty$} & \multicolumn{1}{c}{$\lambda_0$} & \multicolumn{1}{c}{U$_\text{es}$/U$_\infty$} & \multicolumn{1}{c}{St$_0^*$} & \multicolumn{1}{c|}{$f_{0c}$ c/U$_\infty$} & \multicolumn{1}{c}{$\delta$/c} & \multicolumn{1}{c}{$\delta^*$/c}\\
                            \noalign{\smallskip}\hline\noalign{\smallskip}
100-10-P & 4 & --- & 4 & --- & 0.466 & 0.138~c & 1.02 & 0.456 & 3.36 & 0.0754 & 0.0234\\
100-10-F & 3.52 & --- & 3.52 & --- & 0.498 & 0.154~c & 1.02 & 0.488 & 3.23 & 0.0701 & 0.0234\\
 & & & & & & & & \\
150-10-P & 4.14 & 2.76 & 4.14 & 2.55 & 0.497 & 0.125~c & 1.04 & 0.478 & 3.98 & 0.0652 & 0.0285\\
150-10-F & 4.35 & 2.02 & 4.35 & 2.02 & 0.519 & 0.131~c & 1.04 & 0.499 & 3.96 & 0.0607 & 0.0274\\
\noalign{\smallskip}\hline   
\end{tabular}
\normalsize
\caption{Spectra results, Strouhal number and boundary layer quantities.}
\label{Table: Freq. Analysis}
\end{table*}
\begin{eqnarray}
\text{St}_0^* = \frac{f_0 \lambda_0}{\text{U}_{\text{es}}} = \frac{\text{U}_\text{0}}{\text{U}_{\text{es}}}\label{eqn: Strouhal}
\end{eqnarray}

In Figure \ref{Spectra:a}, 150-10-P, the dominant peak is observed at $f_{0-1/2}$, which corresponds to approximately half of the fundamental frequency and indicates the presence of mode 2. Therefore, at this condition, the shear layer upstream and near the trailing edge is already in a non-linear state with pairing of the rollers happening more often. This pairing effect has been seen in many previous studies of planar shear flows \citep{Ho1982, Rajagopalan2005, Rodriguez2013, Perret2009}. It should be noted that this frequency herein is not exactly half of the fundamental peak, which is reasonable as the position of the transition point of the fundamental instability is typically fluctuating in the shear layer; hence this gives a more `broadband' region where this frequency is seen \citep{PRASAD1997, Rajagopalan2005, Dong2006, YARUSEVYCH2009}.

Once the flaplets are attached to the aerofoil and the flow is studied again at the same conditions, Figure \ref{Spectra:d}. The $f_{0-1/2}$ peak is obviously suppressed and the $f_{0}$ is now the dominant frequency. Meanwhile, as shown further below, the flaplets oscillate at their natural frequency with an average amplitude of about 0.4mm (0.2\%c), excited by the observed rollers over the trailing edge. The shift of the dominant peak back to the fundamental instability indicates that the shear-layer has now been stabilised and the growth of non-linear modes has been damped, thus reducing the tendency of vortex pairing.  This stabilisation is thought to be caused by a lock-in effect between the oscillating flaplets and the shear layer fundamental instability, once the frequency of the latter exceeds the natural frequency of the flaplets. Such that they start to oscillate in the flow. This conclusion is based upon a similar observation of flow stabilisation with flaplets attached to the aft part of an aerofoil and a bluff body \citep{Brucker2014, Rosti2017, Kunze2012}, where the flaps act as `pacemaker' and alter the shedding cycle, leading to a reduction in drag and lift fluctuations. It is worth to note that the herein observed stabilisation goes together with a slight reduction in the thickness of the local boundary layer edge ($\delta$). Table \ref{Table: Freq. Analysis}, shows that with the flaplets the thickness has been reduced by 7.5\%, inferring that with the attached flaplets the aerodynamic performance of the aerofoil might also benefit. It is hypothesised that the reduced thickness is the consequence of less pairing events in the stabilised situation, because the pairing causes a strong wall-normal momentum exchange and therefore a thickening of the boundary layer.

For the lowest Reynolds Number cases (100-10-P \& 100-10-F), there was no dominating non-linear instabilities or vortex pairing (i.e. $f_{0-1/2}$) observed in the spectral analysis. The reason for this is thought to be because of the sub-critical state of the shear-layer formation at the lower flow speed \citep{PRASAD1997, Rajagopalan2005}. 

When analysing the velocity probe data for the zero degree cases, it was seen that no obvious spectral peaks were present. The shear layer at these conditions is again expected to be in the sub-critical state as the adverse pressure gradient is weaker compared to the 10$^\text{o}$ angle of attack situation \citep{Huang1995}.

\subsection{Flaplet Motion}
\label{sec:Flap Motion Res.}

\begin{figure}[h]
\centering
{\includegraphics[]{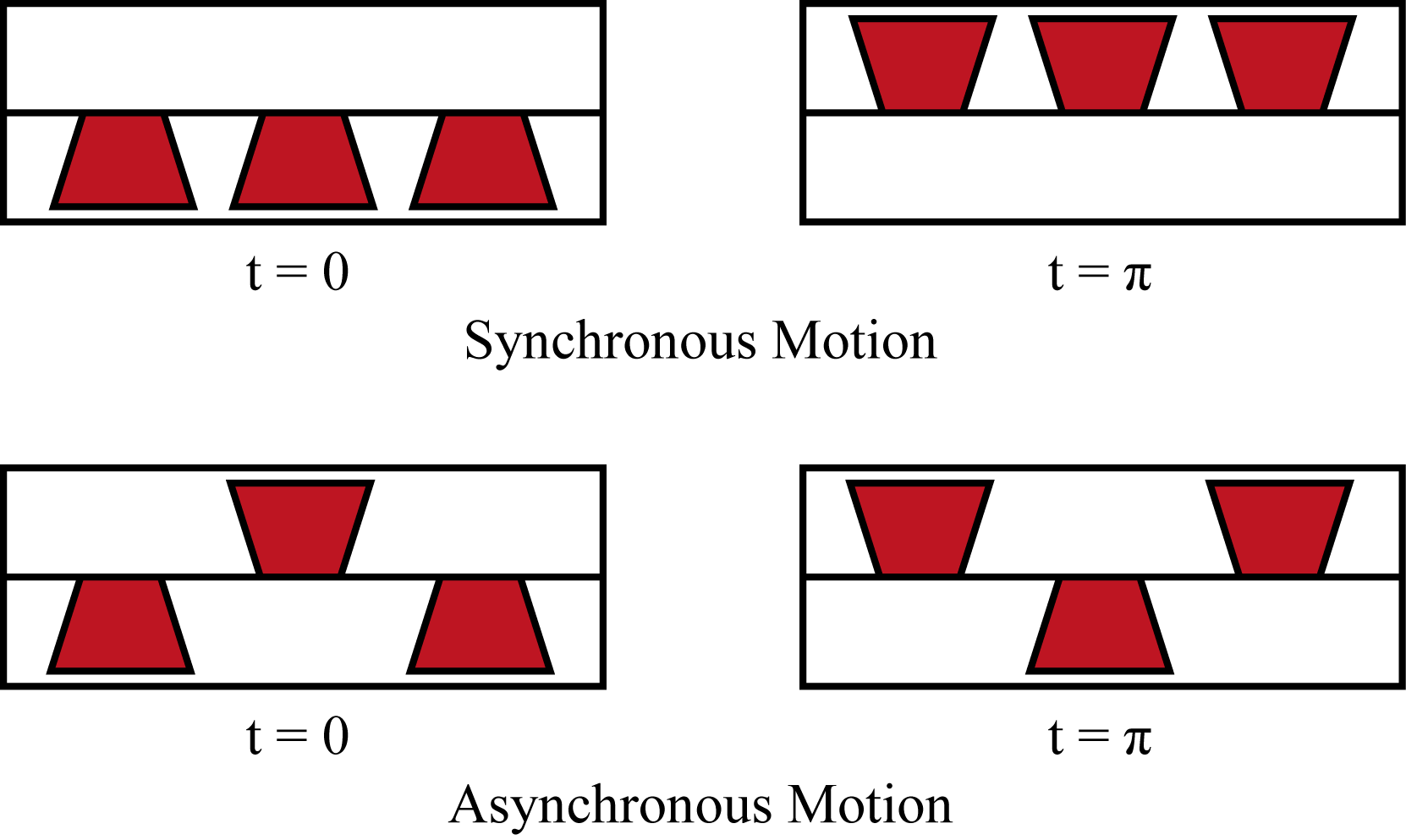}}
\caption{Sketch of the flaplets in either synchronous or asynchronous motion, looking from a downstream view of the trailing-edge.}
\label{fig: Sync/Async}
\end{figure}

For the motion study, three neighbouring flaplets were analysed in order to distinguish between random motion patterns, representing turbulent structures, or spanwise coherent structures, such as the described rollers convecting along the flaplets. Each flap represents a one-sided clamped rectangular beam which oscillates in the first bending mode perpendicular to the long axis, at the natural frequency when being excited. The motion was recorded and analysed for 2.5 seconds leading to a total of 8000 frames. For further analysis, the flaplet motion has been classed into two different categories, synchronous in-phase (S) motion and anti-synchronous (A-S) motion. The S motion is when all three flaplets are in the same phase, both for positive and negative deflections (see Figure \ref{fig: Sync/Async}a for clarity). A-S motion is defined as when the flaplets have a phase delay of $\pi$ with respect to each other (see Figure \ref{fig: Sync/Async}b). A small variance of the phase of $\pm10^\text{o}$ is allowed when the time traces are investigated for such events.  To investigate the probabilities of whether the flaplets have S or A-S motion the natural frequency of the flaplets, $f_1$ = 107~Hz, is used to calculate the maximum number of possible `flap events' in the sampling time, which is 535 over the captured period. Events are only looked for when the center flap is at its maximum positive or negative position. 

\begin{table}[h]
\centering
\begin{tabular}{ccc}
\hline\noalign{\smallskip}
Test Case & P(S) & P(A-S) \\
\noalign{\smallskip}\hline\noalign{\smallskip}
100-0-F      	& 0.104		& 0.023 \\
150-0-F      	& 0.074		& 0.036 \vspace{2mm}\\

100-10-F      	& 0.225		& 0.009 \\
150-10-F	    & 0.162		& 0.016 \\
\noalign{\smallskip}\hline              
\end{tabular}
\caption{Flaplet motion synchronisation probabilities}
\label{Table: Flap Probab}
\end{table}

\begin{figure*}[!ht]
\begin{minipage}{1\linewidth}
\centering
\subfloat[Neighbouring flaplet response when only one flatlet, flaplet 2, is subjected to a delfection]{\label{Coupling:a}\includegraphics[scale=1]{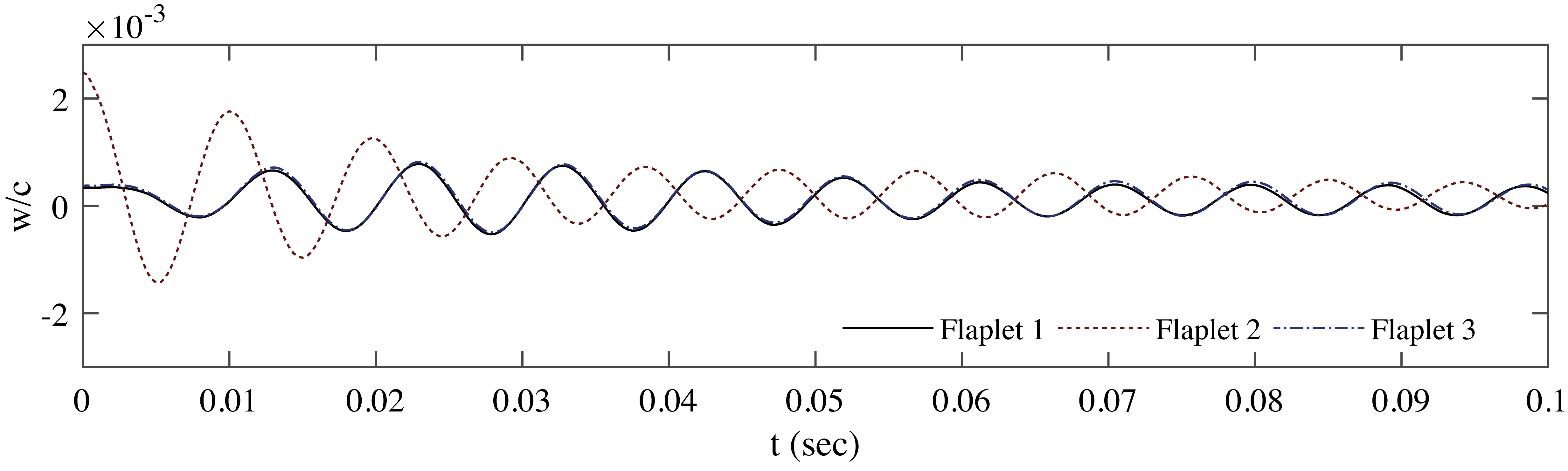}}
\end{minipage}\par \medskip
\begin{minipage}{1\linewidth}
\centering
\subfloat[Neighbouring flaplet motion measured at the test case 150-10-F]{\label{Coupling:b}\includegraphics[scale=1]{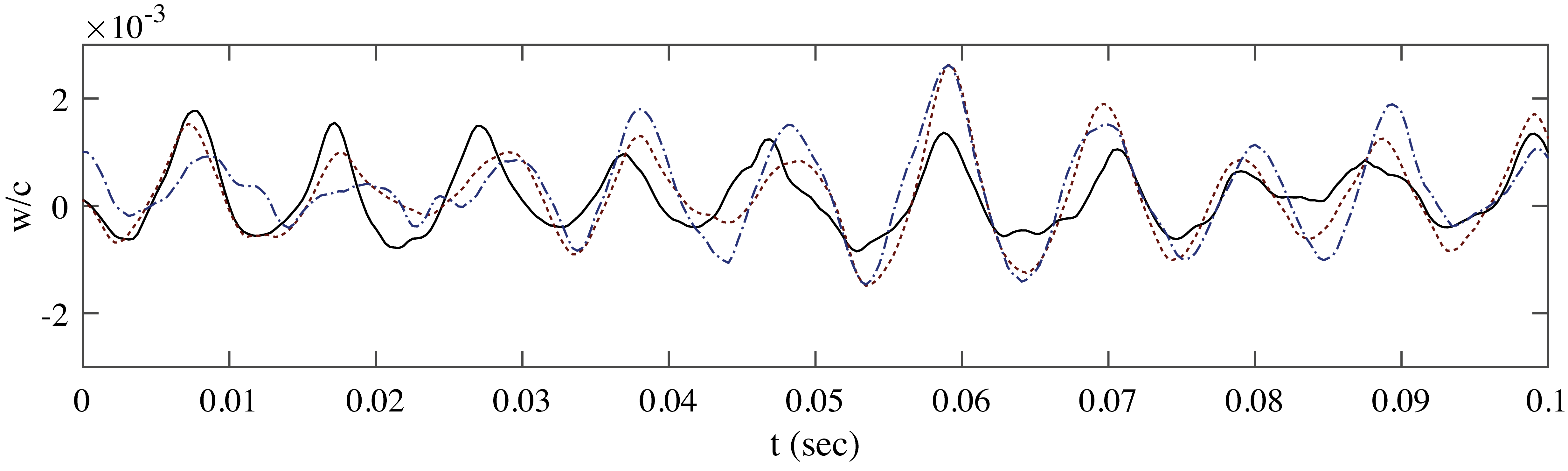}}
\end{minipage}\par
\caption{Neighbouring flaplet motion}
\label{fig: Coupling}
\end{figure*}

When observing the 10$^\text{o}$ cases in Table \ref{Table: Flap Probab}, it can be seen that the S motion has a significantly higher probability than the A-S motion. This is proposed to be due to the presence of strong spanwise coherent shear layer vortices. One might argue that the observed increase is due to the spanwise mechanical coupling of the flaplets. However as can be seen in Figure \ref{Coupling:a}, if one flaplet is excited then the neighbouring flaplets are coupled with a phase delay of $\pi$, as all flaplets are connected via a base layer. Figure \ref{Coupling:b} shows that at the test case 150-10-F, the neighbouring flaplets are in phase with each other and have a strong correlation with each other (minimum 0.716). This indicated that a spanwise structure excited all three flaplets at the same time. An interesting observation is that the 100-10-F case had a higher probability of S motion compared to 150-10-F. As the shear layer fundamental frequency close to that of the natural  frequency of the flaplets, it is hypothesised that the lock-in effect is more pronounced. A similar observation was made by \citet{Rosti2017} and hence a similar effect is proposed here. In comparison, for an angle of 0$^\text{o}$, the flaplet motion is much more random. This is due to the weaker adverse pressure gradient on the upper side of wing, which reduces the mean shear in the inflection point of the boundary layer. Thus, the motion of the flaplets here is thought to be mainly due to small scale turbulent structures convected downstream and exciting the flaplets in a more random manner. 

\section{Conclusion}
\label{sec: Conc.}
Attaching thin flexible flaplets (oscillators) at the trailing edge of an aerofoil - free to oscillate in wall-normal direction -   has been seen to have a profound upstream effect on the boundary layer at moderate angles of attack (10$^\text{o}$); leading to a stabilisation of the fundamental mode growing in the shear-layer on the suction side, along the second half of the wing. For the plain aerofoil, under the given flow conditions, spanwise coherent rollers are formed in the BL in the linear state and then further downstream the growth of non-linear instabilities leads to pairing of those rollers near the trailing edge. The pairing process is clearly suppressed when the flaplets are attached. They started to oscillate at their fundamental frequency with amplitudes of order of 1~mm (0.5\%c), excited by the convection of the rollers over the trailing edge. This observation leads to the conclusion that the stabilisation is due to a resonance or lock-in between the fundamental instability mode and the attached oscillators (flaplets), as this effect is most pronounced when the frequency of the fundamental shear-layer mode is closer to that of the natural frequency of the oscillators, given that the shear-layer is already in its non-linear state. A previous study on bluff body wakes shows a similar stabilisation effect when flexible flaplets, being attached to the aft part, are getting into resonance with the vortex shedding cycle \citep{Kunze2012}.  A further consequence of the lock-in stabilisation is the observed reduction in the boundary layer thickness. This is explained by the reduced probability of pairing events, which otherwise are responsible for larger wall-normal momentum exchange and a further thickening of the boundary layer. It is hypothesised therefore that the flaplets have a beneficial effect on the integral performance of the aerofoil such as reducing drag and increasing lift. This is supported by recent results of active trailing edges oscillators investigated at high frequency \citet{Jodin2018}. The present study also supports on physical means the earlier observation of the flaplets causing the dampening of trailing edge noise in the low-frequency range \citep{Kamps2017}. More recent, \citet{Talboys2018a} have carried out an aeroacoustic study with the present configuration and indeed confirmed the reduction of acoustic noise in the low to medium frequency range (100~Hz - 1~kHz). 

Compared to the very recent study on high-frequency active oscillating trailing edges \citet{Jodin2018}, the present method represent a charming alternative as a passive flow control technique, which does not require sophisticated and costly active manipulation techniques. As shown herein, the flaplets act as self-excited `pacemakers' to stabilise the shear layer on the suction side and therefore it is reasonable to suggest that they also improve the aerodynamic performance of the aerofoil, besides helping to reduce trailing edge noise as already proven in \citet{Talboys2018a}. The natural frequency of these passive oscillators can be tailored for different flow situations either by using different shapes, intelligent materials with temperature or pressure depending properties or by simply applying a deployment strategy where the length of the freely extending tips of the flaplets is varied by sliding the sheet along the trailing edge. 

\section{Acknowledgements}
The position of Professor Christoph Br\"{u}cker is co-funded by BAE SYSTEMS and the Royal Academy of Engineering (Research Chair no. RCSRF1617$\backslash$4$\backslash$11), which is gratefully acknowledged.

%
%


\begin{thebibliography}{32}
\providecommand{\natexlab}[1]{#1}
\providecommand{\url}[1]{{#1}}
\providecommand{\urlprefix}{URL }
\expandafter\ifx\csname urlstyle\endcsname\relax
  \providecommand{\doi}[1]{DOI~\discretionary{}{}{}#1}\else
  \providecommand{\doi}{DOI~\discretionary{}{}{}\begingroup
  \urlstyle{rm}\Url}\fi
\providecommand{\eprint}[2][]{\url{#2}}

\bibitem[{Boutilier and Yarusevych(2012)}]{Boutilier2012}
Boutilier MS, Yarusevych S (2012) {Parametric study of separation and
  transition characteristics over an airfoil at low Reynolds numbers}. Exp
  Fluids 52(6):1491--1506, \doi{10.1007/s00348-012-1270-z}

\bibitem[{Br{\"{u}}cker and Weidner(2014)}]{Brucker2014}
Br{\"{u}}cker C, Weidner C (2014) {Influence of self-adaptive hairy flaps on
  the stall delay of an airfoil in ramp-up motion}. J Fluids Struct 47:31--40,
  \doi{10.1016/j.jfluidstructs.2014.02.014}

\bibitem[{Carruthers et~al.(2007)Carruthers, Thomas, and
  Taylor}]{Carruthers2007}
Carruthers AC, Thomas ALR, Taylor GK (2007) {Automatic aeroelastic devices in
  the wings of a steppe eagle Aquila nipalensis}. J Exp Biol
  210(23):4136--4149, \doi{10.1242/jeb.011197}

\bibitem[{Dong et~al.(2006)Dong, Karniadakis, Ekmekci, and Rockwell}]{Dong2006}
Dong S, Karniadakis GE, Ekmekci A, Rockwell D (2006) {A combined direct
  numerical simulation-particle image velocimetry study of the turbulent near
  wake}. J Fluid Mech 569:185--207, \doi{10.1017/S0022112006002606}

\bibitem[{Dovgal et~al.(1994)Dovgal, Kozlov, and Michalke}]{Dovgal1994}
Dovgal AV, Kozlov VV, Michalke A (1994) {Laminar boundary layer separation:
  Instability and associated phenomena}. Prog Aerosp Sci 30(1):61--94,
  \doi{10.1016/0376-0421(94)90003-5}

\bibitem[{Geyer et~al.(2010)Geyer, Sarradj, and Fritzsche}]{Geyer2010}
Geyer TF, Sarradj E, Fritzsche C (2010) {Measurement of the noise generation at
  the trailing edge of porous airfoils}. Exp Fluids 48(2):291--308,
  \doi{10.1007/s00348-009-0739-x}

\bibitem[{Geyer et~al.(2017)Geyer, Kamps, Sarradj, and
  Br{\"{u}}cker}]{Geyer2017}
Geyer TF, Kamps L, Sarradj E, Br{\"{u}}cker C (2017) {Passive Control of the
  Vortex Shedding Noise of a Cylinder at Low Reynolds Numbers Using Flexible
  Flaps}. 23rd AIAA/CEAS Aeroacoustics Conf pp 1--11, \doi{10.2514/6.2017-3015}

\bibitem[{Gruber et~al.(2010)Gruber, Azarpeyvand, and Joseph}]{Gruber2010}
Gruber M, Azarpeyvand M, Joseph P (2010) {Airfoil trailing edge noise reduction
  by the introduction of sawtooth and slitted trailing edge geometries}. Proc
  20th Int Congr Acoust ICA 10(August):1--9

\bibitem[{Herr and Dobrzynski(2005)}]{Herr2005}
Herr M, Dobrzynski W (2005) {Experimental Investigations in Low-Noise
  Trailing-Edge Design}. AIAA J 46(6):1167--1175, \doi{10.2514/1.11101}

\bibitem[{Ho and Huang(1982)}]{Ho1982}
Ho CM, Huang LS (1982) {Subharmonics and vortex merging in mixing layers}. J
  Fluid Mech 119:443--473, \doi{10.1017/S0022112082001438}

\bibitem[{Howe(1991)}]{Howe1991}
Howe MS (1991) {Aerodynamic noise of a serrated trailing edge}. J Fluids Struct
  5(1):33--45, \doi{10.1016/0889-9746(91)80010-B}

\bibitem[{Huang and Lin(1995)}]{Huang1995}
Huang R, Lin C (1995) {Vortex shedding and shear-layer instability of a
  cantilever wing at low reynolds numbers}. 33rd Aerosp Sci Meet Exhib
  33(8):1398--1403, \doi{10.2514/3.12561}

\bibitem[{Jaworski and Peake(2013)}]{Jaworski2013}
Jaworski JW, Peake N (2013) {Aerodynamic noise from a poroelastic edge with
  implications for the silent flight of owls}. J Fluid Mech 723:456--479,
  \doi{10.1017/jfm.2013.139}

\bibitem[{Jodin et~al.(2018)Jodin, Rouchon, Scheller, and
  Triantafyllou}]{Jodin2018}
Jodin G, Rouchon JF, Scheller J, Triantafyllou M (2018) {Electroactive morphing
  vibrating trailing edge of a cambered wing : PIV , turbulence manipulation
  and velocity effects}. In: IUTAM Symp. Crit. flow Dyn. Involv.
  moving/deformable Struct. with Des. Appl., Santorini, Greece

\bibitem[{Kamps et~al.(2016)Kamps, Geyer, Sarradj, and
  Br{\"{u}}cker}]{Kamps2017a}
Kamps L, Geyer TF, Sarradj E, Br{\"{u}}cker C (2016) {Vortex shedding noise of
  a cylinder with hairy flaps}. J Sound Vib 388:69--84,
  \doi{10.1016/j.jsv.2016.10.039}

\bibitem[{Kamps et~al.(2017)Kamps, Br{\"{u}}cker, Geyer, and
  Sarradj}]{Kamps2017}
Kamps L, Br{\"{u}}cker C, Geyer TF, Sarradj E (2017) {Airfoil Self Noise
  Reduction at Low Reynolds Numbers Using a Passive Flexible Trailing Edge}.
  In: 23rd AIAA/CEAS Aeroacoustics Conf., American Institute of Aeronautics and
  Astronautics, Reston, Virginia, June, pp 1--10, \doi{10.2514/6.2017-3496}

\bibitem[{Kunze and Br{\"{u}}cker(2012)}]{Kunze2012}
Kunze S, Br{\"{u}}cker C (2012) {Control of vortex shedding on a circular
  cylinder using self-adaptive hairy-flaps}. Comptes Rendus - Mec
  340(1-2):41--56, \doi{10.1016/j.crme.2011.11.009}

\bibitem[{Meyer et~al.(2007)Meyer, Cavar, and Pedersen}]{Meyer2007}
Meyer KE, Cavar D, Pedersen JM (2007) {POD as tool for comparison of PIV and
  LES data}. 7th Int Symp Part Image Velocim pp 1--12

\bibitem[{Miklosovic et~al.(2004)Miklosovic, Murray, Howle, and
  Fish}]{Miklosovic2004}
Miklosovic DS, Murray MM, Howle LE, Fish FE (2004) {Leading-edge tubercles
  delay stall on humpback whale ( Megaptera novaeangliae ) flippers}. Phys
  Fluids 16(5):L39--L42, \doi{10.1063/1.1688341}

\bibitem[{Osterberg and Albertani(2017)}]{Osterberg2017}
Osterberg N, Albertani R (2017) {Investigation of self-deploying high-lift
  effectors applied to membrane wings}. Aeronaut J 121(1239):660--679,
  \doi{10.1017/aer.2017.10}

\bibitem[{Perret(2009)}]{Perret2009}
Perret L (2009) {PIV investigation of the shear layer vortices in the near wake
  of a circular cylinder}. Exp Fluids 47(4-5):789--800,
  \doi{10.1007/s00348-009-0665-y}

\bibitem[{Ponitz et~al.(2014)Ponitz, Schmitz, Fischer, Bleckmann, and
  Br{\"{u}}cker}]{Ponitz2014}
Ponitz B, Schmitz A, Fischer D, Bleckmann H, Br{\"{u}}cker C (2014)
  {Diving-flight aerodynamics of a peregrine falcon (Falco peregrinus)}. PLoS
  One 9(2), \doi{10.1371/journal.pone.0086506}

\bibitem[{Prasad and Williamson(1997)}]{PRASAD1997}
Prasad A, Williamson CHK (1997) {The instability of the shear layer separating
  from a bluff body}. J Fluid Mech 333:S0022112096004326,
  \doi{10.1017/S0022112096004326}

\bibitem[{Rajagopalan and Antonia(2005)}]{Rajagopalan2005}
Rajagopalan S, Antonia RA (2005) {Flow around a circular cylinder-structure of
  the near wake shear layer}. Exp Fluids 38(4):393--402,
  \doi{10.1007/s00348-004-0913-0}

\bibitem[{Rodr{\'{i}}guez et~al.(2013)Rodr{\'{i}}guez, Lehmkuhl, Borrell, and
  Oliva}]{Rodriguez2013}
Rodr{\'{i}}guez I, Lehmkuhl O, Borrell R, Oliva A (2013) {Direct numerical
  simulation of a NACA0012 in full stall}. Int J Heat Fluid Flow 43:194--203,
  \doi{10.1016/j.ijheatfluidflow.2013.05.002}

\bibitem[{Rosti et~al.(2017)Rosti, Kamps, Br{\"{u}}cker, Omidyeganeh, and
  Pinelli}]{Rosti2017}
Rosti ME, Kamps L, Br{\"{u}}cker C, Omidyeganeh M, Pinelli A (2017) {The
  PELskin project-part V: towards the control of the flow around aerofoils at
  high angle of attack using a self-activated deployable flap}. Meccanica
  52(8):1811--1824, \doi{10.1007/s11012-016-0524-x}

\bibitem[{Schl\"{u}ter(2010)}]{Schluter2010}
Schl\"{u}ter JU (2010) {Lift Enhancement at Low Reynolds Numbers Using
  Self-Activated Movable Flaps}. J Aircr 47(1):348--351, \doi{10.2514/1.46425}

\bibitem[{Semeraro et~al.(2012)Semeraro, Bellani, and Lundell}]{Semeraro2012}
Semeraro O, Bellani G, Lundell F (2012) {Analysis of time-resolved PIV
  measurements of a confined turbulent jet using POD and Koopman modes}. Exp
  Fluids 53(5):1203--1220, \doi{10.1007/s00348-012-1354-9}

\bibitem[{Stanek(1965)}]{Stanek1965}
Stanek FJ (1965) {Free and Forced Vibrations of Cantilever Beams With Viscous
  Damping}. Tech. Rep. June, National Aeronautics and Space Administration,
  Washington DC

\bibitem[{Talboys et~al.(2018)Talboys, Geyer, and Br{\"{u}}cker}]{Talboys2018a}
Talboys E, Geyer TF, Br{\"{u}}cker C (2018) {The Aerodynamic And Aeroacoustic
  Effect Of Passive High Frequency Oscillating Trailing Edge Flaplets}. In:
  IUTAM Symp. Crit. flow Dyn. Involv. moving/deformable Struct. with Des.
  Appl., Santorini, Greece

\bibitem[{Thomareis and Papadakis(2017)}]{Thomareis2017}
Thomareis N, Papadakis G (2017) {Effect of trailing edge shape on the separated
  flow characteristics around an airfoil at low Reynolds number: A numerical
  study}. Phys Fluids 29(1):014101, \doi{10.1063/1.4973811}

\bibitem[{Yarusevych et~al.(2009)Yarusevych, Sullivan, and
  Kawall}]{YARUSEVYCH2009}
Yarusevych S, Sullivan PE, Kawall JG (2009) {On vortex shedding from an airfoil
  in low-Reynolds-number flows}. J Fluid Mech 632:245,
  \doi{10.1017/S0022112009007058}

\end{thebibliography}
\end{document}